*Review*

# Position Sensitive Gaseous Photomultipliers

**Paolo Martinengo [1], Eugenio Nappi [2] and Vladimir Peskov [1,3]\***

[1] PH Div., CERN, Geneva-23, Switzerland CH-1211
[2] INFN Bari, Bari, Italy
[3] UNAM, Mexico, Mexico

E-Mails: paolo. martinengo@cern.ch; eugenio.nappi@ba.infn.it; vladimir.pskov@cern.ch

\* Author to whom correspondence should be addressed; Tel.: +41-22-767-4643; Fax: +41-22-767-9480 (Vladimir.peskov@cern.ch)

**Abstract:** Advances in the technologies associated with position sensitive gaseous detectors especially featuring CsI as reflective photoconverters will be reviewed. These photodetectors represent the most effective solution for what concerns cost and performance in the case of large area Cherenkov imaging applications in relatively low rate (or low occupancy) high energy physics and astrophysics experiments. Moreover, they are the only choice when the Cherenkov detector is embedded in a magnetic field. Recently proposed single-photon MPGDs (Micro-Pattern-Gaseous Detectors) will be also discussed in view of the successful efforts so far made to extend their sensitivity to visible light. With some modifications, photosensitive gaseous detectors can also be used in the imaging of X-rays and particles.

## 1. Introduction

Nowadays there are three types of position-sensitive photodetectors capable of detecting single photoelectrons: vacuum, gaseous and solid-state. Modern imaging vacuum detectors are: microchannel plates (MCP), position-sensitive multianode photomultipliers and hybrid photodetectors (for details see the review paper [1] and references therein). The main advantages of this type of detector are: high quantum efficiency (20-40% for visible light) and good timing properties (reaching a picoseconds level for some detectors). A typical position resolution of multianode photomultipliers is a few mm, MCP - up to a few μm, and hybrid photomultipliers ~0.1mm. The main disadvantage of vacuum photodetectors is their limited sensitive area: the largest size of multianode photomultipliers is about 5x5cm$^2$, for MCP, it is about 10x10cm2 whereas the largest hybrid photodetector has a cathode diameter of about 20cm.

In the last decade a breakthrough in solid-state detectors technology occurred: Si detectors capable of operating in avalanche mode were developed. The most impressive in this family of devices is called Si photomultiplier (see for example [2]). It is an array of avalanche microphotodiodes manufactured on a common Si substrate. The dimension of each single avalanche photodiodes can vary from 20 to 100μm and their density can be up to 1000 per square millimeters. Each Si photomultiplier (SiPM) pixel operates in Geiger mode. The Geiger discharge is quenched when the voltage on the pixel electrodes decreases below the breakdown value due to external resistors connected to each individual pixel. The gain in Geiger mode is as high as $10^6$ allowing the detection of single photoelectrons Another important feature of this detector is its high quantum efficiency for visible light which is equal or even higher than that of conventional vacuum photomultipliers thus making SiPMs unique for many applications (see for example [3]). However, SiPMs have several disadvantages, for example a large noise level and a high cost, which currently restricts their area of applications.

In this paper we will review a large and important class of photodetectors called Position Sensitive Gaseous Photomultipliesr (PSGPMs). Their main advantage with respect to vacuum and solid- state detectors is the possibility to build detectors with very large sensitive area (up to ~10m$^2$) at a relatively low cost. Most of these detectors today are UV sensitive only. However, with some modifications PSGPMs can be used in the detection of X-rays as well as charged particles. There are also efforts to develop PSGPMs sensitive to visible light. Thus, PSGPMs are quite a large class of detectors widely used in various experiments and applications.

There are two main types of PSGPMs:
1. Ones based on the principle of gas photoionization
and
2. Ones working on the principle of photoelectrons
extraction from solid photocathodes.

## 2. PSGPMs working on the principle of gas photoionization

The first position-sensitive gaseous photodetectors were developed in two parallel works: one made at CERN for RICH application [4] and one in Moscow for plasma applications [5]. These two pioneering works triggered rapid and very successful developments of PSGPMs [5-9]. These first detectors were multiwire proportional chambers (MWPCs) invented earlier by G, Charpak[1] but with the following essential modifications: they had a window transparent for UV and were filled with gases with low ionization potentials (benzene vapors in Ref. [4], sensitive to wavelength λ<135nm and toluene vapors in [5], sensitive to λ<146 nm)—see Fig. 1. UV photons penetrating through the window caused the photoionization of vapors and created primary photoelectrons. These electrons in turn drifted towards the anode wires and in their vicinity where the electric field is very high (typically $10^5$ V/cm) triggered Townsend avalanches. By taking signals from different wires it was possible to determine the 2D position of the avalanche with an accuracy of ~0.1x3 mm$^2$. The latest designs are usually the same MWPCs but combined with large absorption/ drift regions [6, 10]. The main focus, however was not on the design optimization but on finding gases with the smallest possible ionization potential—$E_i$.

---

[1] For the invention of the MWPC G. Charpak was awarded by the Noble prize in Physics in 1992

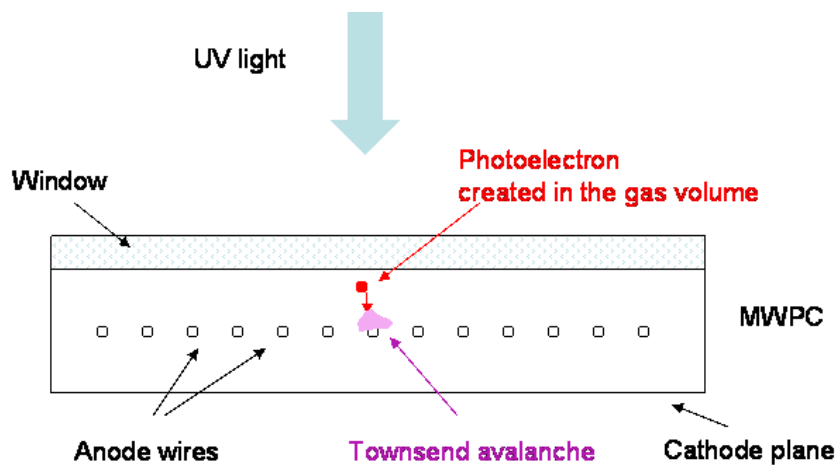

Fig.1. A schematic drawing illustrating the principle of operation of the photosensitive MWPC filled with gases with small $E_i$. In most designs to increase the sensitivity they are combined with large drift/absorption regions (see [10]).

The first successfully used vapors with small ionization potentials were Trimethylamine (TMA) [11] and ethylferrocene (EF) [12]. Later, Triethylamine (TEA) [6] and tetrakis (dimethylamino) ethylene (TMAE) [6, 13] started dominating in the applications. Typical gas gains of photosensitive MWPCs are more than $10^5$, which are sufficient for the detection of single photoelectrons. Fig. 2 shows the quantum efficiencies (QE) of the most commonly used photosensitive gases today: TEA, TMAE, and EF.

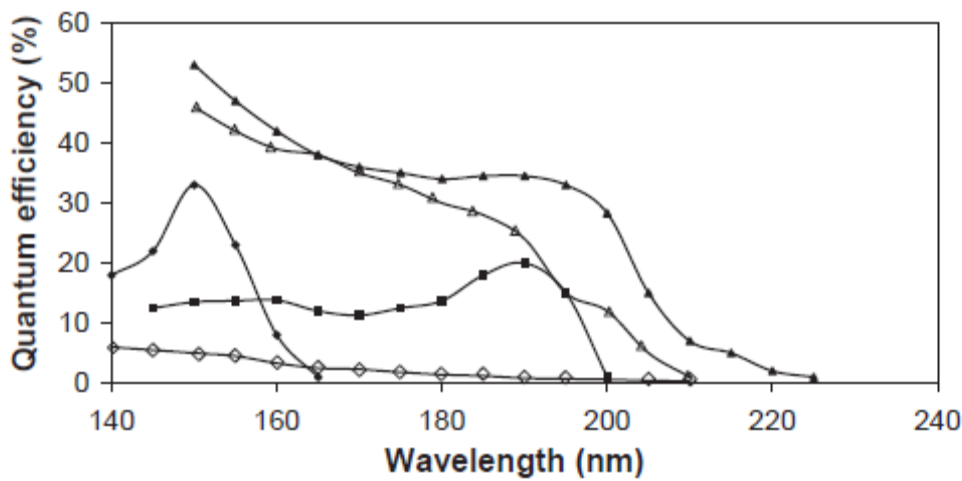

Fig. 2. Quantum efficiency of the most commonly used photosensitive vapours: TMAE (filled triangles), EF (filled squares), and TEA (filled diamonds). For comparison, the QE of the CuI (open diamonds) and CsI (open triangles) solid photocathodes are also shown.

One can see that in some wavelength interval the quantum efficiency is as high as the best vacuum photomultipliers have (above 20%). In contrast to usual vacuum photodetectors, PSGPMs operate at 1atm and thus do not have any mechanical constraints on the window size; this allows them to be built with a very large sensitive area. The other important advantage of gaseous photodetectors is that they are practically insensitive to the magnetic fields which are often used in high-energy physics experiments.

### 3. PSGPMs with solid photocathodes

The threshold of photoionization detector sensitivity is determined by the ionization potential of the gases. As one can see from Fig. 2, the gases with the lowest ionization potential—EF and TMAE, have the ionization threshold at 6.08 eV ($\lambda$ =200 nm) and 5.28 eV ($\lambda$ =230nm), respectively. The EF and TMAE were discovered to be photosensitive elements long time ago and afterwards, there has been no real progress in finding better substances (materials with lower ionization potentials). This is mainly because the substances with low ionization potentials are usually chemically aggressive and/or unstable in air and this is why there have been continuous efforts to replace photosensitive vapors by solid photocathodes.

The advantages of such an approach are:

1) very low thresholds of spectral sensitivity, which is determined by the photosensitive material's work function $\varphi$, 2) excellent time resolution (better than 1 ns) since there is no space jitter in creating photoelectrons; and 3) comparing them to position-sensitive vacuum or solid-state detectors, they are cheaper and can be manufactured with a large sensitive area.

The design of the first wire chamber with a CuI photocathode having a high QE for UV is schematically presented in Fig. 3 [12, 14].

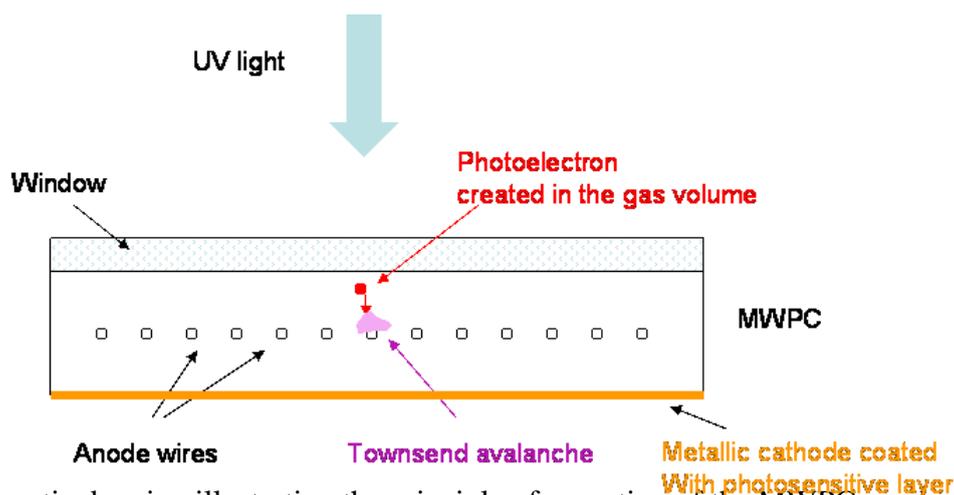

Fig. 3. A schematic drawing illustrating the principle of operation of the MWPC combined with a solid photocathode (metallic photocathode covered with a photosensitive layer, for example CuI or CsI).

In this case, photons penetrating through the window cause photoelectric effect from the solid cathode (no photoionization of the gas). The QE of the CuI photocathode is shown in Fig. 2. Photoelectrons extracted from the cathode move toward the anode wire and triggere Townsend avalanches. Nowadays, PSGPMs with solid photocathodes are considered as very promising new detectors able to replace PSGPMs with photosensitive vapors in many applications. Present developments
can roughly be divided into two categories: 1) UV sensitive detectors with CsI photocathodes and 2) detectors of visible photons.

*3.1. PSGPMs combined with CsI photocathodes*

Nowadays, PSGPMs with CsI photocathodes receive large attention (see paragraph 4.1.1b). The first attempt to combine a CsI photocathode with gaseous detector (a parallel-plate chamber) was illustrated in [15]. CsI was deposited on a metallic plate by a vacuum deposition technique. Surprisingly, the quantum efficiency (QE) of the CsI photocathode remains high enough after it was transferred in air from the evaporation system into the gaseous detector. This simple technology suggested a wide range of application for CsI photocathodes in gaseous detectors. Shortly after that, Dangendorf et al. [16] tested a wire chamber with a CsI photocathode and Seguinot *et al.* [17] performed a systematic study of the CsI photocathode and how to improve its QE. The QE of the CsI photocathode is presented in Fig. 2**.** One can see that it is as high as TMAE. Gains of $(3-5) \times 10^4$ were achieved without any serious feedback problems.
A breakthrough in the application of CsI thin film to Ring Imaging Cherenkov detectors (RICH) detectors was achieved by the RD26 Collaboration at CERN (see paragraph 4.1.1b).

*3.2 Hole- type UV sensitive PSGPMs*

In the previous paragraphs wire-type PSGPM were describes, in these detectors avalanche multiplication of primary photoelectrons occurs near thin anode wires. Applications of PSGPMs were widely extended (see review paper [18]]) when solid photo cathodes, in particular made of CsI, were combined with another gas multiplication structure called "hole- type" [19-21]. In a few words, the modern hole-type detector is a dielectric sheet metalized on both sides in which an array of holes is created by a chemical etching technique or by mechanical drilling (see Fig. 4). If a voltage V is applied across the metalized surfaces, the field lines will exhibit a focusing effect inside the holes, as shown in Fig. 4**,** and a very high electric field can be created in this region so that at some critical value of V (V>Vc) the avalanche multiplication of primary electrons can occur.

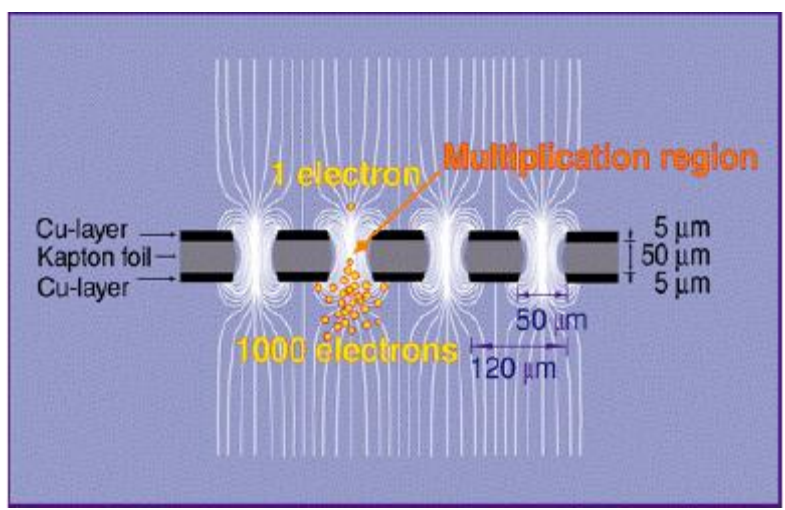

Fig.4. The principle of operation of the hole-type gaseous detectors: once a voltage drop is applied between the electrodes a strong electric filed is created inside the holes. If a primary electron enters the hole and the voltage V is above a critical value Vc, a Townsend avalanche develops

Hole-type detectors have several advantages over the traditional wire–type detectors. The most important of them are:
1) efficient suppression of ion and photon feedback; this allows the hole–type detector combined with solid photocathode to be operated at high gas gains even in badly quenched gases
2) the possibility of a charge extraction: avalanche electrons created by the multiplication processscan be extracted from the capillary's holes and directed to towards a second multiplication structure or to a 2-D readout plate

First hole-type multiplication structures successfully combined with high efficient photocathodes were a glass capillary plate [22] and the so called "Gas Electron Multiplier" or GEM [23]. The glass capillary tubes are commercially available and are a by-product of the manufacturing of the microchannel plate (MCPs). Typically, such plates have a diameter of 20 mm, a thickness of 0.8 mm, and a hole diameter of 100μm (see Fig. 5). Unfortunately capillary plates are quite expensive, consequently the main focus on further developments was on GEMs.

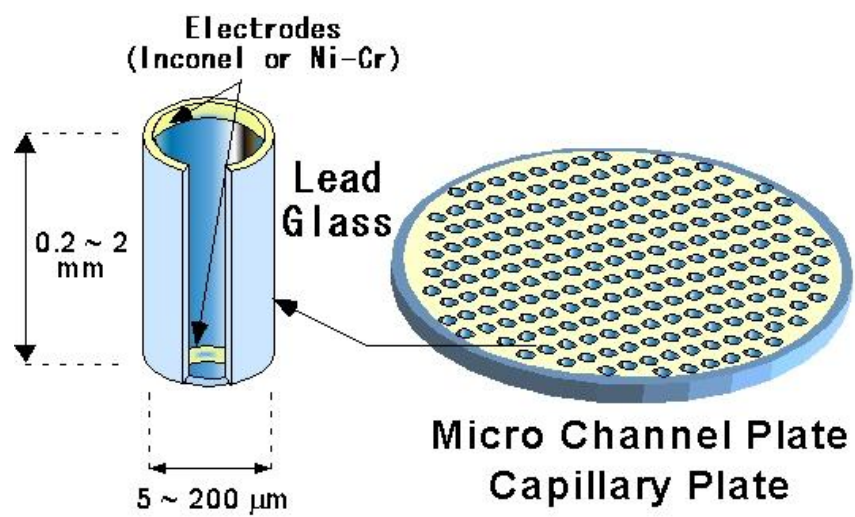

Fig. 5. A schematic drawing of glass capillary plate [20] used in designs of first gaseous photomultipliers sensitive to visible light [22].

GEM is a hole-type structure in which the dielectric sheet is manufactured from a 50 μm thick Kapton. This important modification allowed the production of large-area detectors of various shapes. CsI coated GEMs were developed by the Breskin group [24] and further developments of this type of photodetector was done by the PHENIX collaboration [25].

However, GEMs are rather fragile devices: they require dust free conditions during assembling and can be easily damaged by sparks which appear at high gain operation. Unfortunately, occasional sparks are almost unavoidable at high gains. Performed studies show that the maximum achievable gain of hole- type detectors increases with their thickness [26]. This is why our first attempt was to develop a thick GEM (TGEM) [27]. This was a metallized from both sides printed circuit board with drilled holes (Fig. 6). This simple device allows the achievement of a maximum gain of $10^5$ which is 10 times higher than that achievable with the conventional GEM [28, 29]. Nowadays TGEMs with active area of $60 \times 60$ cm$^2$ can be industrially produced [30].

Recently, even a more robust version of the TGEM was developed having electrodes made of high resistivity materials. It was called Resistive Electrodes TGEM or RETGEM. The RETGEM is in fact a hybrid between a Resistive Plate chamber [31, 32] and a hole-type gaseous multiplier. If the voltage is applied to the electrodes, due to their non infinite resistivity, they will charge up and start acting as equipotential surfaces thus creating the same field lines focusing effect inside the holes as in the case of metallic electrodes. However in contrast to ordinary TGEMs they are fully spark-protected: due to the electrodes resistivity the spark energy is reduced by 100-1000 times.

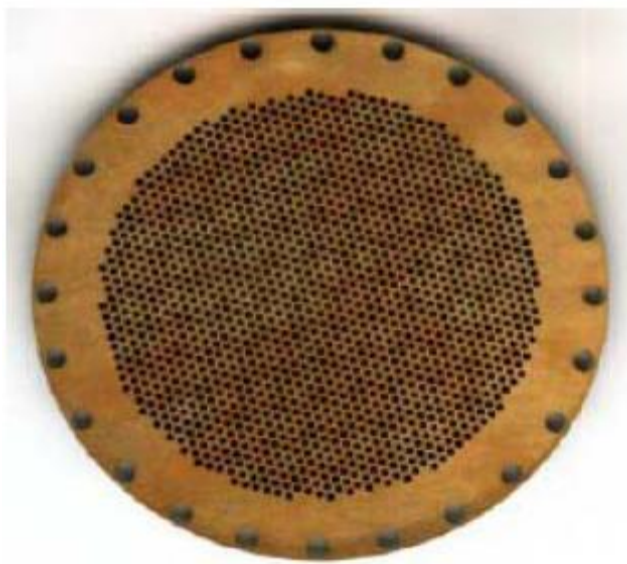

Fig.6.Photo of the first prototype of TGEM manufactured by hole drilling technique [27]

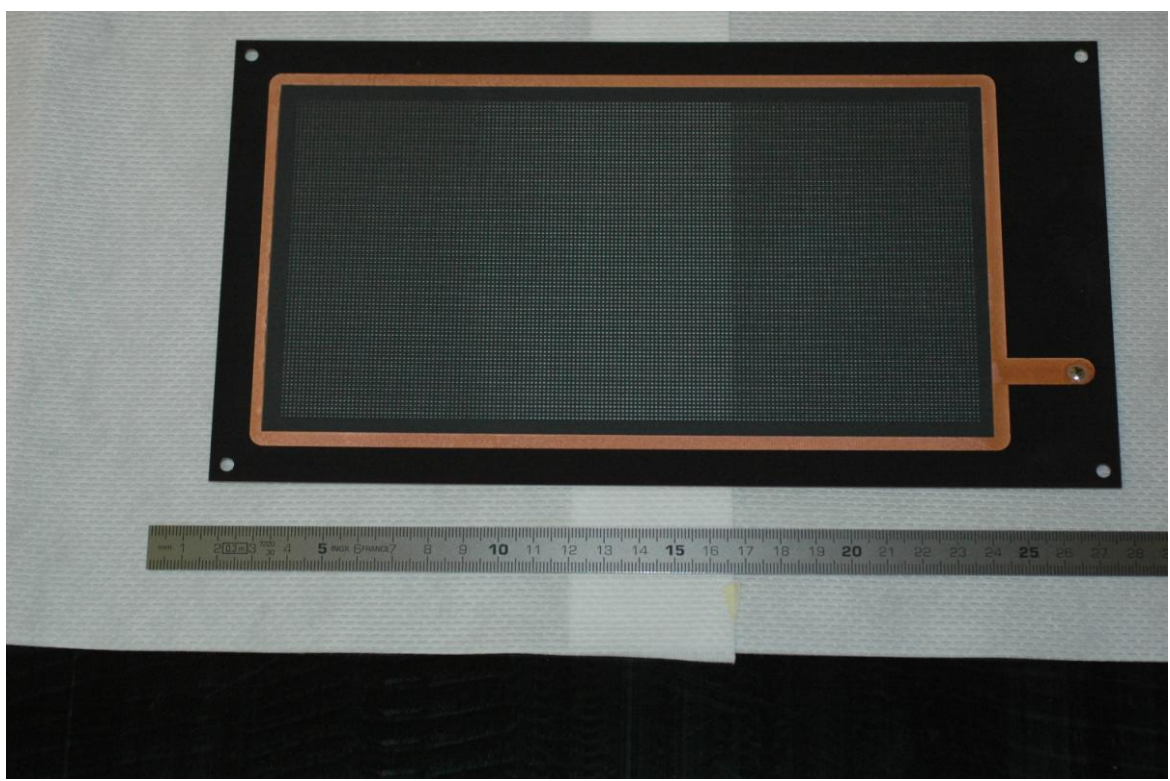

**Fig. 7** Picture of a RETGEM having an active area of 10x20cm$^2$.

Applications of TGEM and RETGEM based gaseous photomultipliers will be described in paragraph 4.2.

*3.3. Gaseous detectors with solid photocathodes sensitive to visible light*

During the last decade or so there have been a lot of efforts in the development of gaseous detectors sensitive to visible light (see review papers [33, 34].

The experience of several groups have however shown that it is not an easy task to employ solid photocathodes sensitive to visible light (SbCs, GaAs/Cs or bialkaline) in an ordinary gas amplification structure, such as wire chambers or parallel-plate chambers (PPAC) [22,35]

There are two main problems:

1) any tiny trace of impurities (for example, oxygen or water) cause degradation of the photocathode quantum efficiency due the photocathode destruction

2) in presence of high quantum efficiency photocathodes, it is almost impossible to reach gains above 10-100 due to the photon and ion feedback.

Both problems actually are linked to the low work function of the photocathode. Therefore, in the case of photocathodes sensitive to visible light, they represent a serious drawback.

What are the ways to overcome these problems? The first problem has been solved by using a properly clean gas system.

An alternative approach could be to protect the photocathode by a thin layer of CsI or other material evaporated on top of the photocathode [36]. Such a layer should be thin enough to be transparent to both photons and photoelectrons created by the photocathode, but should also slow down the gas diffusion. Preliminary results indicate that the photocathodes with the protective layers have a lower QE but are more robust and have better ageing properties (Fig. 8)**.** This allows the use of less clean gases and even the exposure of the photocathodes to air [36].

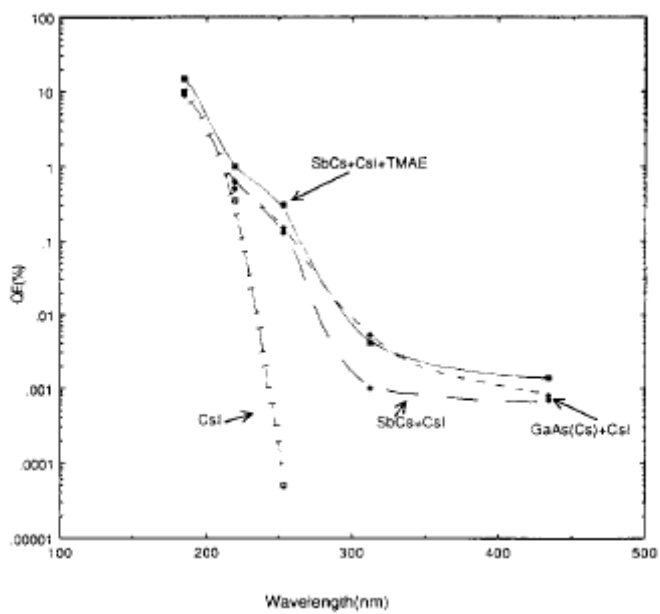

Fig. 8. The quantum efficiency of GaAs/Cs and SbSc photocathodes covered by a 30nm thick CsI protective layer [36]

The work on the photocathode protection was successfully carried by Breskin's group [37]. They confirmed that the best protective layer is the CsI and were able to manufacture quite robust photocathodes which preserved their quantum efficiency even being exposed for 5-10 minutes to oxygen or water vapours (see Fig. 9)

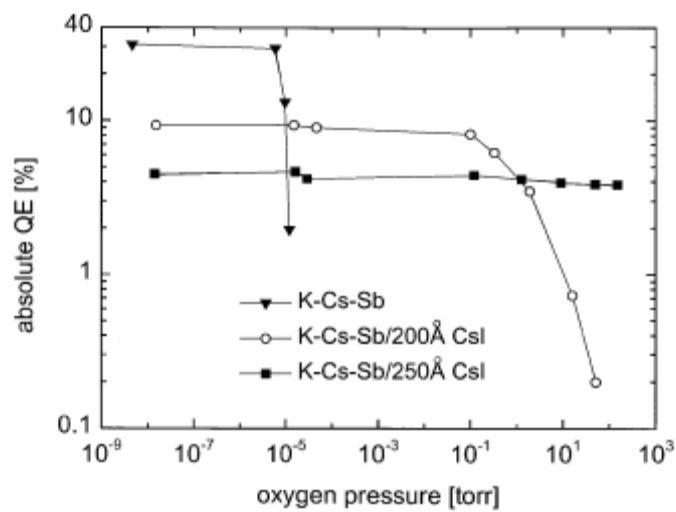

Fig. 9. The evolution of the absolute quantum efficiency of K-Cs-Sb photocathodes expose to oxygen. Shown are the results of uncoated and coated photocathodes with 20nm and 25nm thick CsI protective films, as a function of the residual oxygen pressure. Each point represents 5 min exposure to oxygen, followed by quantum efficiency measurements in vacuum (from [37]).

The second problem (photon and ion feedback, which spoils the performance and also causes discharges) is more complicated to fix. The discharges in gaseous detectors due to feedback appear when one of the following conditions is fulfilled first: $Ak\gamma_{ph}=1$ or $Ab\gamma_{+}=1$, where A is the gas gain, $\gamma_{ph}$ and $\gamma_{+}$ are the probabilities to create a secondary electron from the cathode due to the photoelectric effect or ion interactions respectively and coefficients k and b describes what fraction of photon and ions from the avalanches are reaching the cathode.

In ordinary gaseous detectors (wire–type or parallel-plate type with metallic electrodes) $\gamma_{ph}$ and $\gamma_{+}$ $<10^{-6}$ and the coefficients k and b are equal to unity. However, in the case of highly efficient photocathodes, sensitive to visible light, $\gamma \sim 10^{-2}$. Therefore, the maximum achievable gains cannot be more than $10^2$ (as was actually observed experimentally). To reach higher gains, special designs allowing suppression of photon and ion feedbacks should be developed. These developments acquired a new momentum with hole- type micropattern detectors (capillaries and GEM). The hole-type multipliers have important advantages over the traditional avalanche detectors due to the efficient reduction or suppression of photon and ion feedbacks (due to the avalanche confinement in the holes), thus $k<<1$ and $b<1$. As was already mentioned earlier, first successful attempt was done with glass capillary plates-see Fig. 10 and [22]. Glass is a material which is perfectly compatible with vacuum technologies and conditions and it meets all requirements for the photocathode manufacturing.

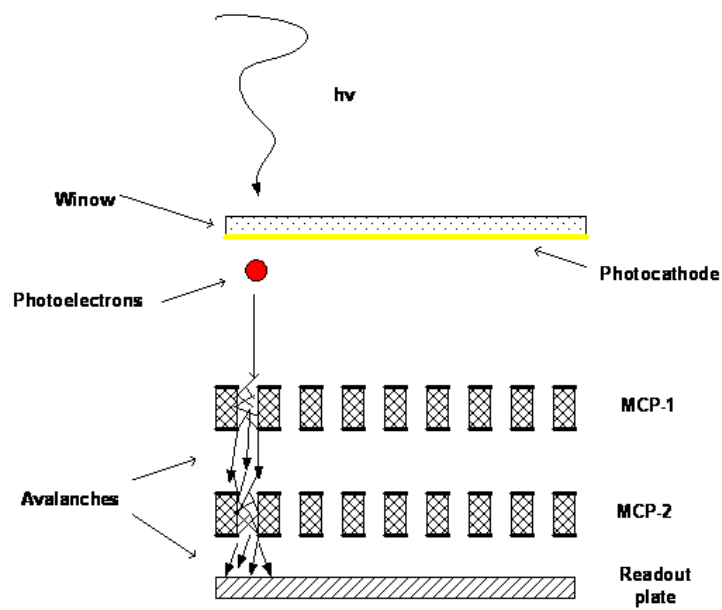

Fig.10. A schematic drawing of the first gaseous photomultiplier sensitive to visible light [22]. It is based on a double capillary plate operating in cascade mode.

Several prototypes of gaseous photomultipliers, based on capillary plates, were built and successfully tested (see for example [38, 39]), allowing detection of visible light with an efficiency of a few percent.

These studies triggered other developments. For example, Hamamatsu Inc. produced several commercial prototypes of gaseous detectors based on capillary plates combined with photocathodes sensitive to visible light (see Fig. 11)

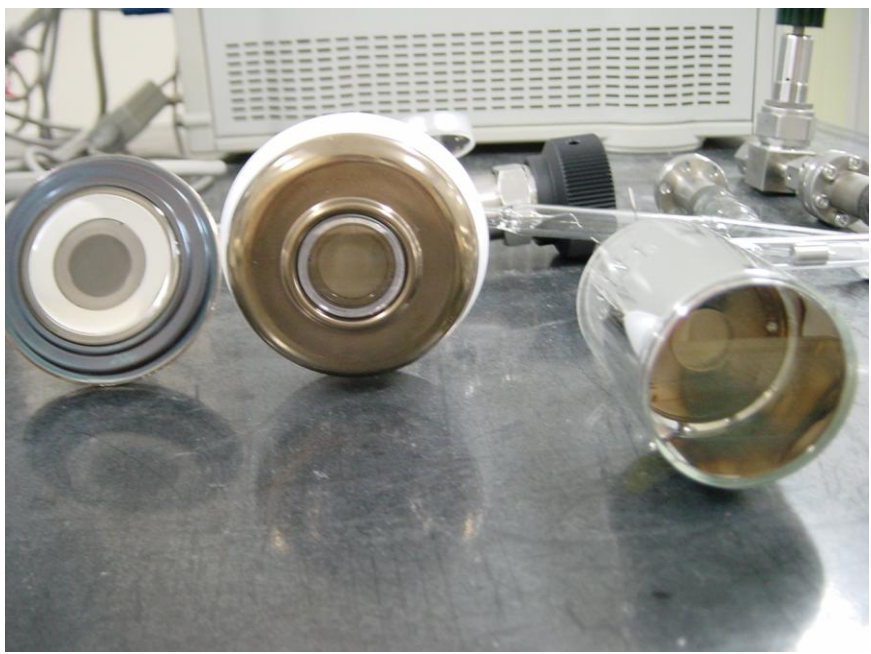

Fig.11. Commercial prototypes of capillary based gaseous photomultipliers [40].

A great success was recently achieved by Breskin's group who developed a state of the art high efficiency gaseous photomultiplier based on a special design of GEM detectors [41]-see Fig. 12.

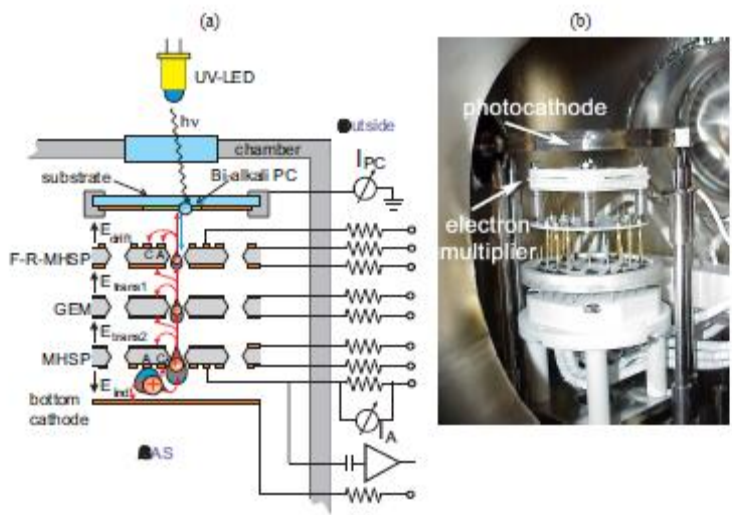

Fig. 12. High efficiency gaseous photomultiplier developed in [41]: a) a schematic drawing of a triple GEM with special pattern electrodes allowing a strong suppression of ion feedback, b) a photograph of this detector

## 4. Applications

*4.1 "Traditional" applications*

PSGPMs have much different application. Certainly, Ring Chrenkov Counters (RICH) is the main application of PSGPMs today. Several large-scale photosensitive detectors were build for high-energy physics and astrophysics experiments.

*4.1.1. Ring Imaging Cherenkov detectors: Earlier ideas and first designs*

In [42] it was proposed a Cherenkov counter capable to detect individual Cherenkov photons and the pattern formed by them. This was a big progress in the field because the existing detectors were able only to discriminate particles emitting Cherenkov light versus others below the Cherenkov threshold.

This novel approach is an evolution of the concept described in [43] in which the focusing of Cherenkov photons emitted along the particle's track is obtained by employing a reflective mirror (of radius R and mirror focal length F=R/2) thus obtaining a circular ring image of radius r in the mirror focal plane . In the small angle approximation

$$r = F \tan \theta_c, \qquad (1)$$

where $\theta_c$ is the so called Cherenkov angle, i.e. the angle between the emitted Cherenkov radiation and the particle path. This is related to the particle's velocity v by

$$\cos \theta_c = c / nv \qquad (2)$$

where *c* is the speed of light and *n* is the refractive index of the medium.

The schematic layout of the RICH detector proposed in [42] is shown in Fig.13. In this RICH design, a spherical photosensitive detector is placed at a distance of R/2 from the inner mirror surface. As a possible photon detector in this paper an array of needle detectors was considered (see Fig.14). It was proposed to be filled with one of the following photosensitive vapours: benzene ($E_i$=9.15eV), Cis-2-butene( $E_i$=9.35eV) or acetone ($E_i$=9,65).

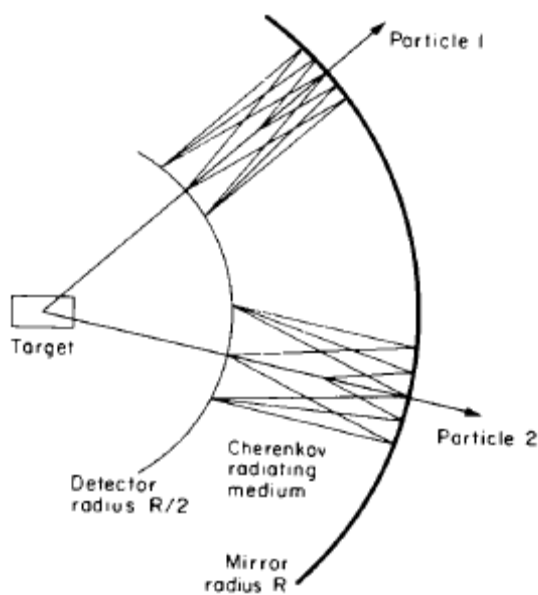

Fig.13. Schematic drawing of the RICH detector proposed by J. Seguinot and T. Ypsilantis. The ring images produced by different particles emerging from the target are shown (from [42]).

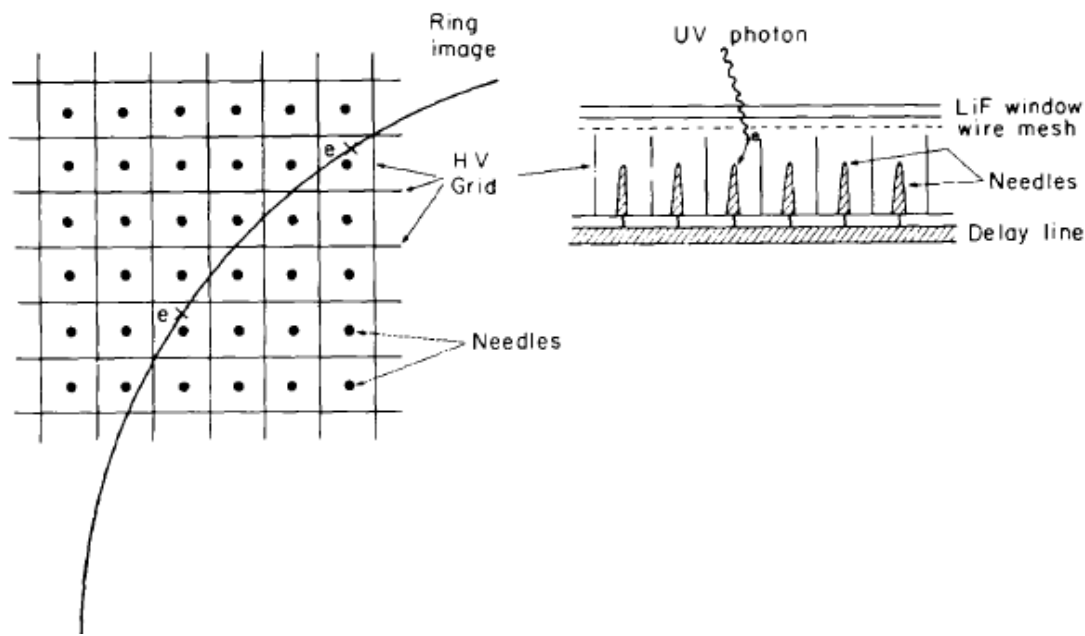

Fig. 14. The photosensitive detector proposed in [42] for the detection of Cherenkov rings: an array of needle type Geiger- or- proportional counters filled with vapours having small $E_i^*$

Another RICH geometry, the so called proximity focusing RICH detectors, again combined with photosensitive gases, was tested by Charpak et al in 1979 [45] (Fig.15). In the proximity-focusing design, a thin radiator volume emits a cone of Cherenkov light which traverses a small distance - the proximity gap - and is detected on the photon detector plane. The image is a ring of light with a radius defined by the Cherenkov emission angle and the thickness of the proximity gap. The ring thickness is mainly determined by the thickness of the radiator (and chromatic aberration). In work [45] the proximity focusing ring was detected by a multistep avalanche chamber followed by a spark chamber. After the photon conversion, the light emitted by the sparks was recorded by a photocamera.

---

* Note that a similar design of photodetector was develop and used for plasma studies [44, 7]

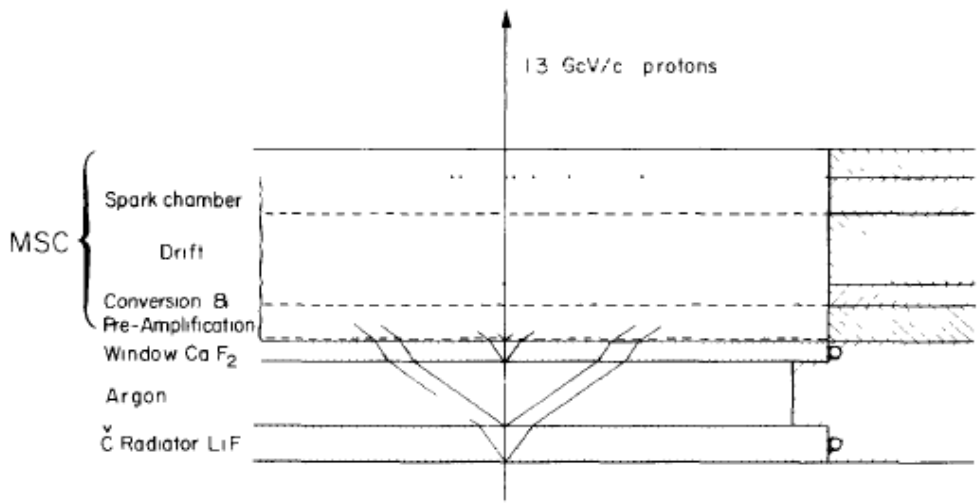

Fig.15. The multistep avalanche chamber used in [45] for imaging the photons emitted by the Cherenkov effect in a 5mm thick LiF radiator (a proximity focusing RICH). The chamber window was a 3 mm thick CaF$_2$ crystal. The ring image was formed directly in the UV-sensitive detector with a dispersion that depended on the radiator thickness

*4.1.2. Present status: RICH detectors based on photosensitive MWPCs*

Considerable advances in the technologies associated with MWPC photon detector manufacturing have recently extended the potentialities of the devices based on the measurement of the Cherenkov angle via the direct imaging of the emitted photons. This well-established technique is largely employed in high-energy and astro-particle physics experiments to achieve the identification of charged particles in an impressively vast momentum range from few hundreds MeV/c up to several hundreds GeV/c.

Gaseous photon detectors represent the most effective solution for what concerns cost and performance in the case of a large area Cherenkov imaging application in relatively low rate (or low occupancy) experiments. Moreover, they are the only choice when the Cherenkov detector is embedded in a magnetic field (recall that gaseous detectors can operate at rather high magnetic fields without degradation of their performance such as position resolution).

The centroid measurement of the charge induced on the cathode plane provides an accurate, two-dimensional localization of the photon conversion point, allowing the reconstruction of overlapping photon patterns from a multi-particle event.

At moderate amplification gain ($\leq 10^5$), single-electron pulse-height distribution has an exponential shape. In fact in the case of low electric field, the electron ionization is built up with several independent collisions with the gas atoms and therefore the probability *P(q)* that an avalanche has a charge *q* is obtained by the Furry distribution

$$\mathbf{P(q)} = e^{-q/\bar{q}} / \bar{q}$$

(3)

where $\bar{q}$ is the mean charge of the avalanche.

The single-electron detection efficiency is therefore given by

$$\varepsilon = \int_{q_{th}}^{\infty} P(q)dq = e^{-q_{th}/\bar{q}} \qquad (4)$$

where $q_{th}$ is the threshold charge needed to remove the detector noise.

The exponential form of $\varepsilon$ is an unfavorable feature of gas detectors operated at low gains. In fact, a small decrease in the gas amplification implies a strong loss of efficiency. Accordingly, the front-end electronics (FEE) must feature a low noise performance in order to achieve high single photoelectron detection efficiency. In addition, the exponential shape of the single-photoelectron response also implies that a high dynamic range is desirable in order to preserve the spatial resolution allowed by charge weighting.

A more favorable peaked pulse-height distribution, called the Polya distribution, occurs for higher gain values allowing a more stable setting of the electronic threshold.

However, as drawback, at higher gains the gas photon detector experiences a photon feedback caused by photons emitted by the de-excitation of gas molecules after the avalanche mechanism has occurred. Consequently, anode wires must be surrounded by complicated electrode blinds that prevent secondary photons to initiate new avalanches after being converted by the photosensitive agent in the chamber.

In summary, chamber operation at low gas gain is preferable because, in addition to the negligible photon feedback, it has also invaluable advantages on the detector engineering. In fact it reduces the wire ageing rate and the probability of sparking thus allowing to design devices with relaxed mechanical tolerances readily achievable on large surfaces.

4.1.2a. TEA and TMAE -based MWPCs for RICH devices

Up to ten years ago, vapors of TMAE or TEA, added to a regular gas mixture and flushed through the detector volume, represented the only possible choice for gaseous RICH detectors. However, both of them show serious drawbacks when implemented in photon detection systems, although the long experience acquired so far (especially in the case of TMAE) allows safe and reliable handling and utilization. Indeed TMAE causes strong anode wire ageing, its low vapour pressure requires detector operation at high temperature and quite complicated electrode geometry with blind structures to prevent spurious avalanches from feedback photons.

Although TEA is a less reactive chemical compound, it has a higher photo-ionization threshold than TMAE and therefore the choice of radiator media is limited and the use of fragile and expensive far-UV windows becomes mandatory. Moreover, in the operational bandwidth of TEA-based detectors, water and oxygen contents become relevant and materials are strongly chromatic.

Quite many large–scale high energy physics and astrophysics experiments used RICH detectors filled with TMAE or TEA vapors (see for example [6]).

4.1.1b. CsI-based MWPC for RICH

In the quest for new photosensitive materials for gaseous photon detectors, a thin film (few hundreds nm) of CsI deposited onto the cathode plane of a MWPC plays nowadays the role of the best alternative to TMAE and TEA because of its an extremely small photon absorption length and long electron escape length thus allowing for a better time resolution and detector operation at room temperature.

Basic research into the properties of CsI started almost twenty years ago with the publication of the seminal paper by J. Seguinot and T. Ypsilantis [17] who opened up a new direction in the development of gaseous RICH counters proving that QE of reflective CsI photocathodes in $CH_4$ is as high as the QE of CsI in vacuum.

The quantum efficiency (QE) of CsI is the largest among alkali halides used in reflective mode. Moreover thin CsI films can be easily implemented and stored in almost all gases as long as they are free of humidity in contrast to other efficient solid photon converters that are chemically so reactive to be exploitable only in high vacuum devices like PMTs.

Nowadays in the field of RICH applications several experiments such as ALICE, STAR, COMPASS and HADES have inherited the approach pioneered by the RD26 collaboration, led by F. Piuz, based on a modular design consisting of CsI- MWPCs operated in $CH_4$ at normal temperature and pressure.

For example, the ALICE implementation is shown schematically in Fig. 16. It consists of a liquid Cherenkov radiator combined with a MWPC having a pad readout cathode plane segmented into pads coated with a CsI layer.

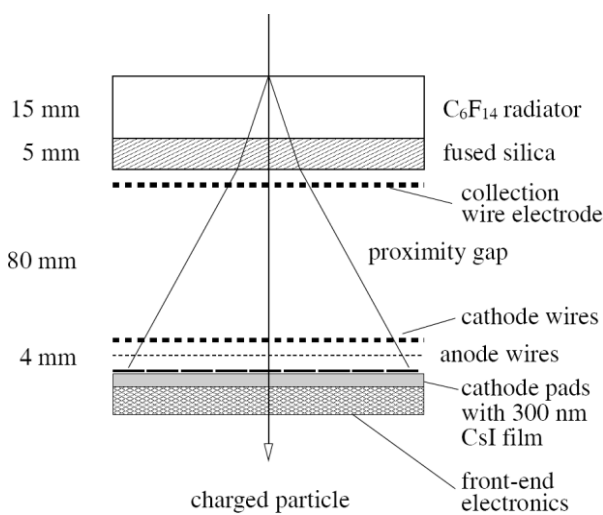

Fig.16. Scheme of principle of the proximity focusing ALICE CsI-RICH detector

As an example in Fig. 17 is shown a Cherenkov ring recorder with one of the ALICE RICH modules

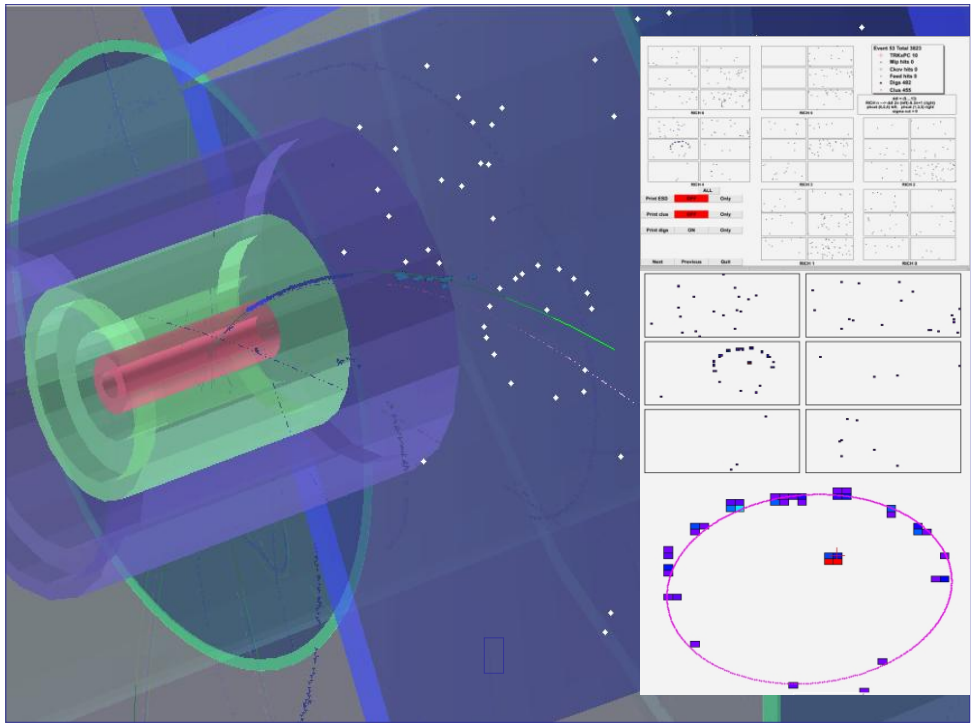

Fig. 17. Example of Cherenkov rings produced by proton-proton collisions at ALICE RICH detector

4.1.1c. GEM-based RICH

The first CsI coated GEM RICH was built for the PHENIX experiment [25]; the schematic drawing of the PHENIX Cherenkov detector is shown in Fig. 18.

The detector consists of a 50 cm long radiator filled with pure $CF_4$, directly coupled to a windowless configuration of a triple GEM detector.

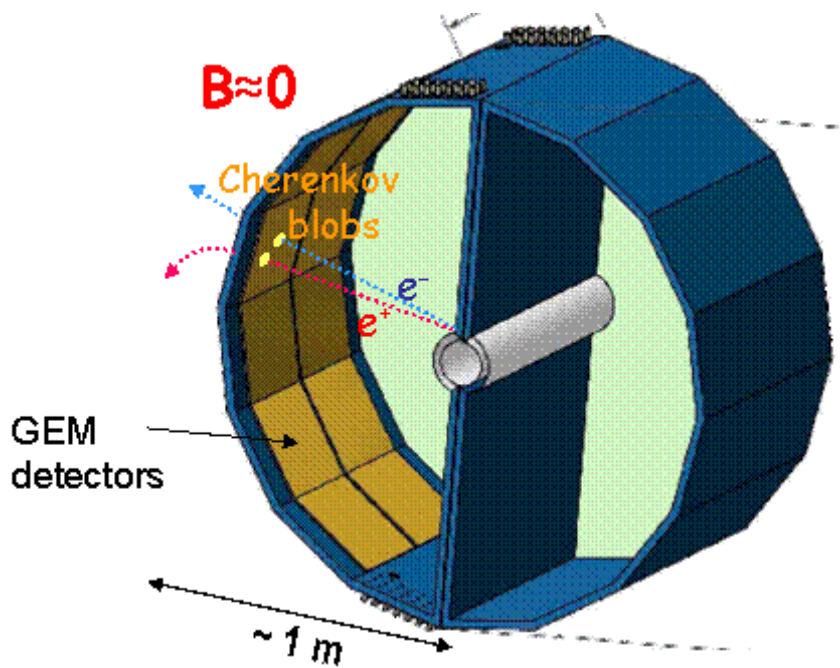

Fig.18. An artistic view of the PHENIX hadron-blind detector based on photosensitive GEMs

The top face of the first GEM layer facing the gas radiator is coated with a thin CsI film. The photoelectrons extracted from the CsI by the Cherenkov light are pulled into the holes of the GEM where they experience an avalanche multiplication. The optimal electric field $E_d$ in the drift region is $E_d \approx 0$ (see Fig. 19). This allows operating this detector at slightly negative $E_d$ and therefore to suppress the signal from charged particles (for this reason the detector is called "hadron blind"-it is in fact ionization blind)-see Fig.20.

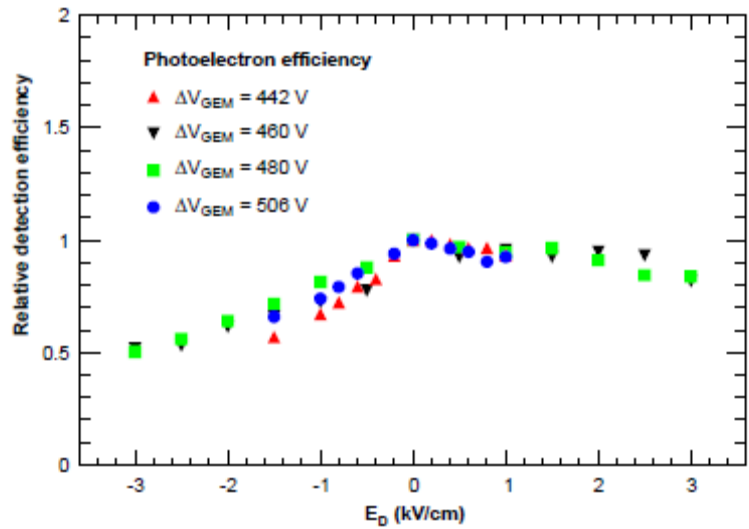

Fig. 19. The photoelectron collection efficiency for various voltages applied across the GEM vs. the electric field Ed in the gap between the drift mesh and the upper GEM (from [25]).

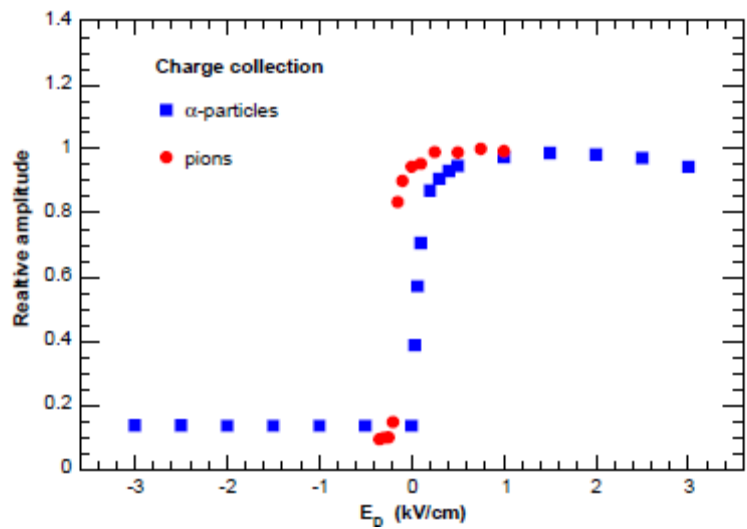

Fig. 20. Collection of ionization charge vs. the drift field Ed in the gap between the mesh and the upper GEM (from [25]).

4.1.1d. Promising new developments: RICH detectors bases on TGEMs and RETGEMs

In the last few years several groups started considering the use of TGEM and RETGEMs for RICH applications [46-48]. In example Fig. 21 a schematic drawing of a small RICH prototype equipped with $CaF_2$ window and with CsI coated triple TGEM or RETGEM photodetectors is shown.

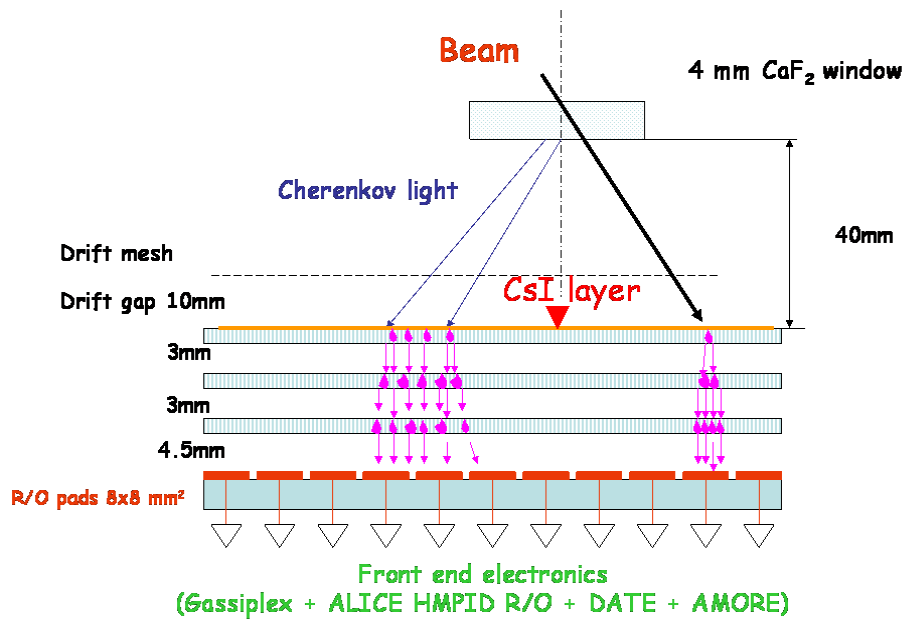

Fig. 21. A schematic drawing of a RICH prototype with a triple CsI coated RETGEM [46].

Preliminary tests supported by simulations indicated that with such detector one can achieve the same efficiency for Cherenkov light as with MWPC. However, TGEMs and RETGEMs offer several important advantages, the most important among them are:

1) they can operate without any feedback problems at 10 time higher gain than the CsI-MWPC

2) they can operate in larger variety of gases, including badly quenched gases.

Fig. 22 shows an image of the beam (red spot) and the Cherenkov light obtained with such a detector during the beam test at the T10 facility at CERN [46]. The COMPASS group also obtained similar results [47].

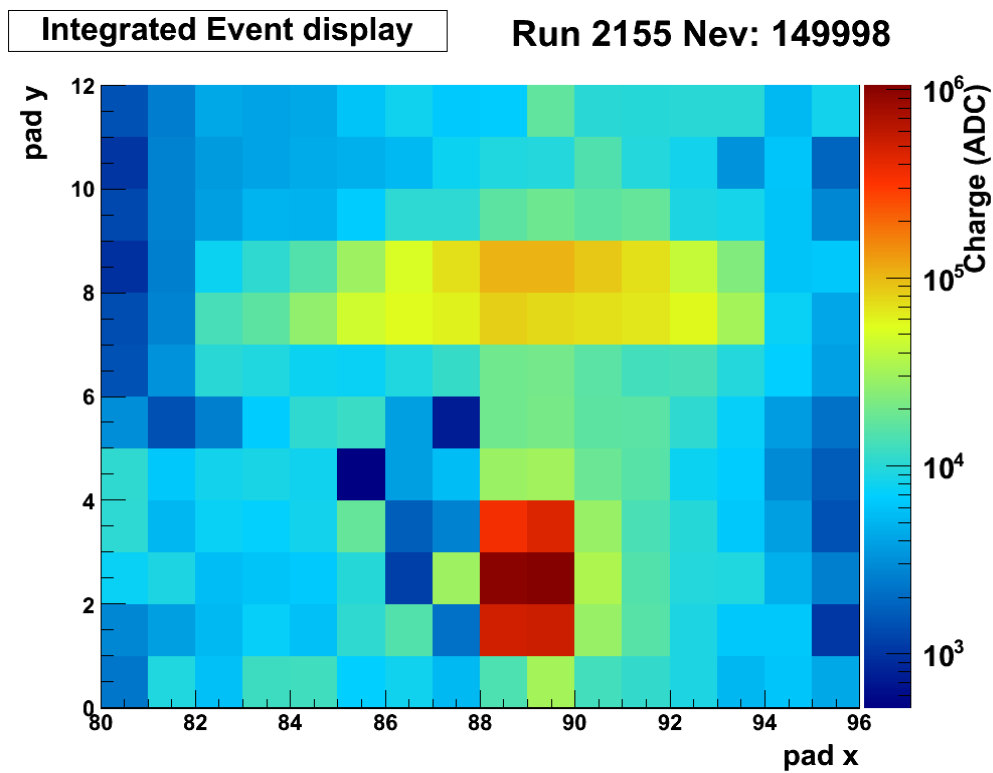

Fig.22. An image of the integrated events obtained during tests of RICH prototype based on CsI coated RETGEMs. The red spot at the bottom of the figure is the image of the particle beam and the yellow strip on the top of the figure is the image of the Cherenkov light. Due to the very large Cherenkov angle and the relatively small size of the detector only a portion, a band, of the ring is detected.

4.1.1e. Other "traditional" applications

Other applications are less massive but clearly demonstrate the growing interest in these types of detectors and their capabilities. Let us shortly comment on these new perspective applications:

1. Plasma diagnostics. Various types of PSGPMs were used for plasma diagnostics: for the visualization of plasma instabilities and as a position sensitive detector in a focal plane of vacuum spectrographs [5, 7-9, 12]

2. Readout of scintillators. It was experimentally demonstrated that PSGPMs are very competitive detectors for the readout of scintillators emitting in UV. Very successful results were obtained with gaseous scintillators (see for example and references therein [11, 13, 16, 27, 33] and crystal scintillators (see for example [49, 50].

3. X-ray, γ and particle detection. In this case solid photocathodes were used as efficient converters. One of the latest developments is a high position resolution X-ray scanner for radiography [51].

*4.2. New areas of application of gaseous imaging detectors.*

In the previous paragraphs we gave a short review of traditional applications of position-sensitive gaseous detectors such as high energy physics and astrophysics (RICH detectors), plasma diagnostics, detectors of x-ray and gamma photons (with possible future application in medicine).

One can expect that in the near feature the application areas of gaseous imaging detectors will greatly expand. The new opportunities may come, for example from micropattern gaseous detectors and especially from recently developed micropattern detectors with resistive electrodes: they are spark- protective and thus more robust and reliable. Below we will give just three examples of new "untraditional" applications borned in the past several years.

4.2.1.Cryogenic photodetectors

There is arising interest now to the so-called "double -phase noble liquid dark matter detectors." It is basically a noble liquid TPC housing two phases (liquid-gas or solid-gas) of a noble element in a single cell. The operational principle is as follows: any interaction inside the volume of the noble liquid will create an ionization track (with $n_0$ primary electrons). Some fraction of the ions $n_0\eta_r$ will recombine and produce a flush of scintillation light which is detected by vacuum photomultipliers surrounding this volume. Because this detector is operating as a Time Projection Chamber (TPCs) with applied drift electric field some fraction of ions and electrons $n_0\eta_d$ will escape recombination and free electrons will drift along the field lines towards the border between the liquid and gaseous phase which they reach after the time $t_d$. As was first shown in work [52] in a rather strong electric filed applied across the border (~10kV/cm) these electrons can be extracted into the gas phase where they can be detected using gaseous detectors. Thus such TPC first record the scintillation light produced by the interaction of the radiation or particles inside the liquid (proportional to $n_0\eta_r$) and then, after the drift time $t_d$, it can detect the charge signal (proportional to $n_0\eta_d$). The ratio of the light to chare signal depends on the nature of the interaction. For example, in the case of recoil tracks with high density of ionization (for example neutrons or WIMPS[*]) $\eta_r >> \eta_d$; whereas in the case of gamma radiation with low density tracks $\eta_r \leq \eta_d$. Hence, by measuring the light to charge ratio one can discriminate between various interactions and select only desirable events. In ordinary noble liquids TPC (for example such as used by ICARUS collaboration in the Gran Sasso laboratory) the drifting electrons are collected on electrodes placed inside the liquid and the minimum amount of the charge which can be measured by this technique is determined by the noise of the electronics. In the case of detection of electrons in the gas phase (as it is implemented in the double-phase detectors) and by exploiting amplification features of the gas detectors the sensitivity can be considerably increased. This method allows combination of good detecting properties of a liquefied gas (a high density and the ability of ionization electrons to drift under the action of an applied electric field) with the potentialities of gas position detectors. Currently most of dark matter noble liquid detectors, for example, ZEPLIN, XENON, WARP (see review paper [53]) are using gas scintillation chambers in the gas phase and the primary and the secondary scintillation lights are detected with vacuum photomultipliers or with solid photodetectors (avalanche diodes or SiPMs). However there are some new developments aiming to use of gaseous photodetectors instead of expensive detectors mentioned above (see, for example [54, 29] and references therein. For example, at present, several groups are considering the use of hole-type gaseous multipliers (GEMs, THGEMs) for the detection of the UV light and primary electrons produced by recoils in noble liquid dark matter detectors (see for example [55] and references therein). Most of early studied were focused to demonstrate that with cascaded GEMs operating in cooled noble gases one can detect charges produced by X-rays in these gases or extracted from the noble liquids (see [55] and references therein). The aim of our recent series of works (see for example [57–60] and references therein) was to investigate if hole-type gaseous multipliers coated with CsI layer also operate stably at cryogenic temperatures and can be used for the detection of scintillation light. It was demonstrated that CsI coated GEMs, capillary plates as well as THGEMs have high quantum efficiencies at cryogenic temperatures for the UV photons and thus can replace expensive photomultipliers usually used in cryogenic TPCs for the detection of the scintillation light. Recently the Novosibirsk group confirmed our results obtained with CsI coated GEMs [61] using a double phase noble liquid detector (see Fig. 23).

---

[*] WIMPs or weakly interacting massive particles **,** are hypothetical particles serving as one possible solution to the <u>dark matter</u> problem.

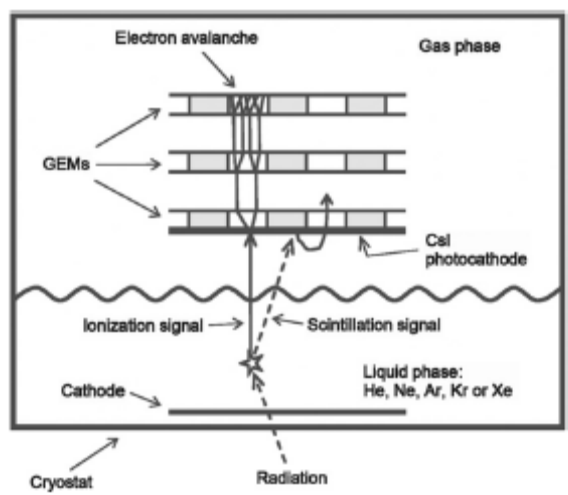

Fig. 23. A schematic drawing of the double phase noble liquid detector with triple photosensitive GEM used in [61].

4.2.2. Detection of flames and dangerous gases with imaging gaseous detectors: recent development

Yearly fires kill thousands of people, destroy properties, and even industrial and agricultural infrastructures. The most violent and dangerous are forest and bush fires which due to their enormous scale can even make an impact of the environment and climate change.

It is obvious that the most effective way to fight fires, including forest fires, is the early identification with automatic monitoring systems. Such system should not only have a high sensitivity, but also a capability to distinguish between the appearance of small fires and various false signals (welding, lightening and cetera). Nowadays, due to the complexity of identification of flames on their early stage, a combination of several detectors is frequently used to reject false alarms, for example, a smoke detector, an infrared and a UV detector integrated in one monitoring system. The most sensitive commercial UV flame detectors, so–called "EN 54-10 class- 1" can detect a ~30x30x30cm$^3$ flame at a distance of ~20m in 20sec. The reason why such a device can detect flames is the following: the sunlight in the wavelength interval 185-280 nm is fully absorbed in the higher part of the atmosphere by the ozone layer, while the atmosphere is rather transparent for these wavelengths at the ground level. To be more precise: it is transparent in the wavelength interval of 185-280 nm for distances of <100 m and transparent for 250-280 nm for longer distances (~km). On the other hand, all flames in air emit in the wavelength interval 185-260 nm. This offers a unique possibility to detect flames and fires without background from the visible and UV light from the Sun and from the visible light in a room. Note that the sensitivity of the class-1 UV detectors is comparable to the sensitivity of the best commercial IR detectors [62].

Studies performed in [59, 63- 65] show that photosensitive gaseous detectors (operating in photon counting mode), depending on their designs, can be 100-1000 times more sensitive to the UV flame emission than the best commercial detectors and thus are very attractive for this application. For example, measurements show that photosensitive gaseous detectors are able to reliably detect a flame of ~1.5x1.5x1.5 m$^3$ at a distance of about ~1km [66]. Because gaseous detectors have a very fast response time (time resolution typically is better than 1μs) they can also promptly record an appearance of various sparks. In combination with compact pulsed UV sources the photosensitive gaseous detectors will also be able to detect smoke and dangerous gases [63, 67].

Having a flame detector with a long operation range, insensitive to day light and prone to false alarms would mean a significant competitive advantage on the fire safety market.

Especially "powerful" are imaging gaseous detectors: these can determine the position of flames (see Fig.24) and be programmed such that most of the false signals are rejected.

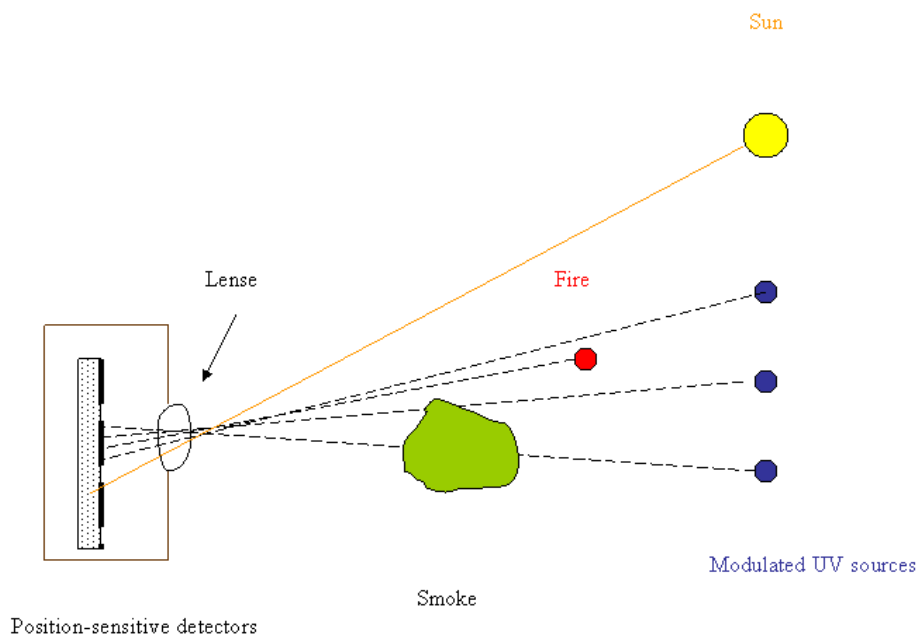

Fig. 24. A schematic drawing illustrating the operation of a position–sensitive gaseous detector combined with a lens in the flames safety application. Imaging capability makes the detector "intelligent": it can detect the position of flame and sparks (red circle in the drawing), reject various false signals and being combined with compact UV pulses source (blue spots) to determine from absorption measurements the appearance of smoke or dangerous gases.

As an example in Fig. 25 is shown a digital image (a number of counts measures from the readout strips) of a small gasoline flame located 300m away from the position-sensitive gaseous detector. Recall that commercial UV flame detectors are at least 100 less sensitive and do not have any imaging capability.

Due to the high sensitivity, good timing characteristics and the position resolution the new detector can be used to survey large area and thus can replace in a cost efficient way several conventional detectors performing the same task.

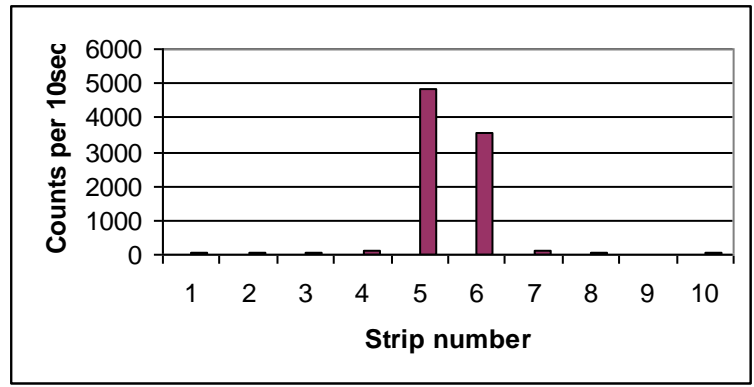

Fig. 25. A digital image of the gasoline flame ~20x20x20cm$^3$ recorded with a gaseous imaging detector based on resistive Thick GEM at a distance of R=300m (from [66]).This detector has a sensitivity more than 100 times higher than the best commercial UV flame detectors.

4.2.3. Hyperspectroscopy

Beside the flame detection there are several other applications which require the detection of weak UV radiation in daylight conditions. One of them is hyperspectroscopy- a new method of surface imaging with simultaneously high position and spectral resolutions ($\leq$ 1nm). This is achieved by a special spectrograph (called a hyperspecrtograph - see Fig. 26) combined with an optical system which allows for monitoring a selected narrow strip (with a length A and the width $\Delta$B) on the surface under study (let's call it a "strip of interest").

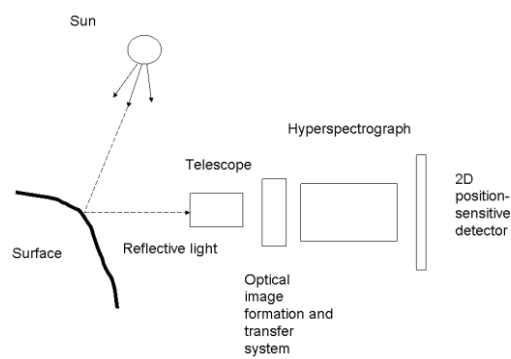

Fig .26. A schematic drawing of a device for hyperspectroscopic imaging. An optical image formation and transfer system allows one to select a strip of "interest" on the investigated surface. The spectrometer forms a 2-D image in its focal plane: in one coordinate a 1D image of the strip in the given spectral interval and in the other coordinate - its spectra.

The hyperspectrograph forms in its focal plane a reduced image of the strip with the size of (A/M)x(ΔB/M), where M is a coefficient determined by the given optical system for each wavelength within the investigated spectral interval. For simplicity let us assume that the two-dimensional (2-D) position resolution which the optical system offers is ~ΔB and the input slit of the hyperspectrograph is adjusted to a value of ΔB/M. In this case an image of the strip with a spectral resolution of ΔS= S ΔB/M, where S is the spectral resolving power of the spectrograph S=Δλ/ΔL (L being the length of the region in the focal plane of the spectrometer on which the given spectra is projected) is formed in the focal plane of the hypespectrograph. To record this image one has to use a 2-D position sensitive photodetector is required (see Fig.26) with a position resolution of 50-100μm. Usually a hyperspectrograph is installed on a flying object, for example on a helicopter or a plane and this allows one to perform a scan of the earth's surface; the selected strip on the surface will move synchronically with the flying carrier. In such a way a two dimensional image of the surface simultaneously with high position and spectral resolutions is obtained. This gives an enormous recognition power since such a method provides much more information than the usual colored pictures or those obtained from direct observations by human eye, for example it allows to make some quantative conclusions about chemical compositions of the surface under survey. Thus the main function of hyperspectral remote sensing image data is to discriminate, classify, identify as well as quantify materials presented in the image. Another important application is the subpixel object detection, which allows the detection objects of interest with sizes smaller than the pixel resolution (ΔB/M in the case of our particular example), and abundance estimation, which allows to evaluate the concentrations of the different signature spectra (compounds) present in the image.

Although the main application of the hyperspectroscopy is the analysis of earth surfaces from helicopters and satellites for geological and environmental purposes such as for the search of spills from oils pipes (see Fig. 27 [68]), in the past few years the areas of application of this method started to expand to other areas, like industry and home security.

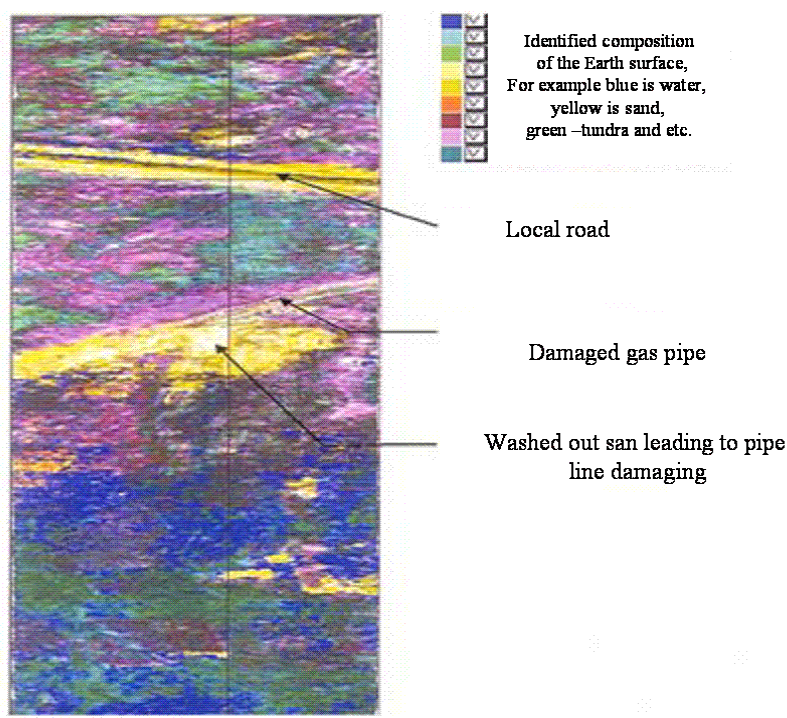

Fig. 27. A hyperspectroscopic computer synthesized image of the earth surface (from [68]. It is possible to identify the surface composition, e.g. the yellow color is sand, the rose color is wet earth, blue is the water and the light blue represents the swam. One can see the sand regions expelled by ground water to the surface together with the gas pipe

It is very attractive to extend the hyperspectroscopic method to the UV region of spectra: 185-280 nm. This offers new technical possibilities in achieving materials and pattern recognitions; for example, it allows to make unique measurements using artificial UV light sources installed in industrial laboratories or even on the helicopters. The main problem in such an approach is that the extremely low intensity of the scattered and reflected UV light from the rather weak artificial source should be detected on the very strong background of the long wavelengths (> 280 nm) produced by the sunlight. Preliminary tests show that photosensitive micropattern gaseous detectors can be an attractive option for this task: they may have large sensitive area, they are practically insensitive to visible light and have very low rate of noise pulses compare to other detectors such as MCP*. One of the first prototypes of such UV hyperspectrograph containing a parallel-plate micropattern detector in its focal plane is described in [69, 70] and is shown in Fig. 28.

---

* One should note that in the last few years there was a significant progress in developing low noise UV sensitive MCPs, for example with CsTe photocathodes, however typically their effective area is about a few cm$^2$ which is insufficient for applications requiring UV visualization in daylight conditions. Moreover in contrast to photosensitive gaseous detectors their sensitivity to the Sun light is not negligible

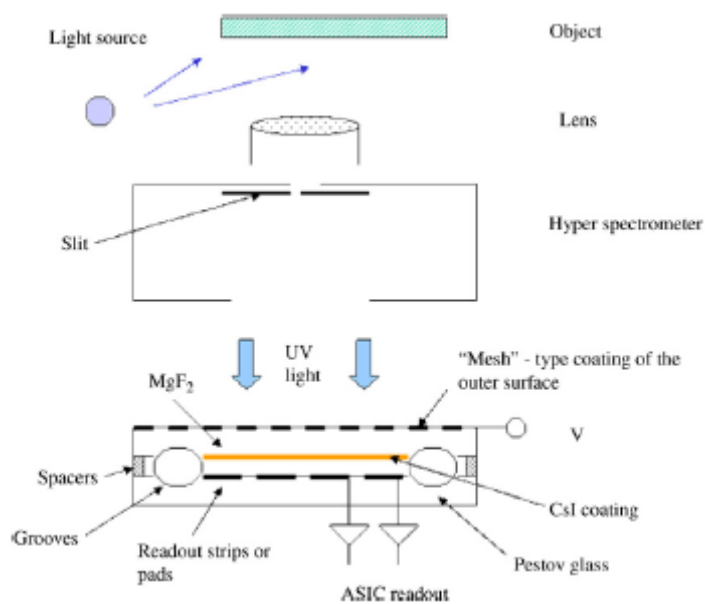

Fig.28. A schematic drawing of the hyperspectrometer [69] combined with a photosensitive a micropattern gaseous detector described in [71].

## 5. Conclusion

As one can see from this short review, PSGPMs are new very promising imaging detectors. Due to their high efficiency, good position and time resolutions they can compete with other kind of imaging detectors, especially in applications requiring large sensitive areas.

## Acknowledgements

Authors would like to thank P. Mezzomo for this help in preparation of this paper

.